\documentclass[10pt,latex8]{article}

\usepackage{latex8}
\usepackage{times}
\usepackage{graphicx}
\begin{document}
\title{An Unified Definition of Data Mining}
\author{Christoph Schommer\\
        University of Luxembourg, Campus Kirchberg\\
        Dept. of Computer Science and Communication\\
        6, Rue Coudenhove-Kalergi, L-1359 Luxembourg\\
        Email: christoph.schommer@uni.lu
}

\maketitle
\thispagestyle{empty}

\begin{abstract}
Since many years, theoretical concepts of Data Mining have been developed and improved. Data Mining has become applied to many academic and industrial situations, and recently, soundings of public opinion about privacy have been carried out. However, a consistent and standardized definition is still missing, and the initial explanation given by \cite{frawley} has pragmatically often changed over the years. Furthermore, alternative terms like \textit{Knowledge Discovery} have been conjured and forged, and a necessity of a \textit{Data Warehouse} has been endeavoured to persuade the users. In this work, we pick up current definitions and introduce an unified definition that covers existing attempted explanations. For this, we appeal to the natural original of chemical states of aggregation.
\end{abstract}

\Section{Evolutionary History}

The idea of exploration of facts exists since thousands of years, since things like criminal cases or the search for the right medical diagnosis can certainly be interpreted as derivatives: in both cases, numerous facts like evidences or symptoms exist, but the understanding of their interplay and togetherness is initially not obvious. Not till then, when a model can be arranged or more precisely, hypothesised, an understanding of the underlying situation becomes likely. Sometimes, external signals support a verification, as for example when the alleged criminal admits the crime or an observation of other patients' health care uniquely leads to the solely possibility; but in almost all situations, a concrete understanding can not be achieved without puzzling. 

Today, the computer machine has found its way into our daily life. Especially in industry and science, an existence without this machine is no longer conceivable as the automatic processing of \textit{data} - which henceforth represents a basic existence - outclasses the human performance in many situations. However, there exist situations in which human power is still the ne plus ultra, but for situations in which the essential capacity goes beyond the human's grasp, the computer machine is indispensable. In this situation, the long existing concept of exploring data gathers momentum, having computer machines finally be the catalyst for puzzling. Furthermore, industrial and scientific persons of different kind have got aware of these automatic possibilities, especially in research fields like \textit{Statistics} or \textit{Machine Learning}. Formerly, living in a cheerless existence while only concentrating on a low number of data, these representatives have got awaken, understanding future challenges and chances.

With that, the term \textit{Data Mining} has been devised allusion to the digging for gold in digital data mines. This figurative illustration has been chosen to get the idea across to users. As a consequence, the deducing of \textit{Data Mining} has been done, representing the symbiosis of modern technology and engineering mechanics. Nowadays, \textit{Data Mining} produces lots of interest, many people launch into it, often taking as it is but sometimes also distorting the basic idea, postulating new research fields in conjunction with industrial applications. One example is \textit{Knowledge Discovery}: to discover knowledge is still not realisable as the complexity of what knowledge is and how we automatically may find it is still out of sight. The coming machine age does not really motivate something else, but the term is still in use and without exception communicated without reflecting. Another example is the invention of a \textit{Data Warehouse}: the fiction that an exploration of data is still impossible without having such a data store, has been accepted for many years in industry although a little cogitation would have revealed that this is nothing else than a marketing strategy. Furthermore, it was told that \textit{Data Mining} is so easy to perform, being a children's game, quasi a standard for almost everyone, without that we can not stay any longer.  However, the reality looks much more different. Today, the users are confused to such extent, that even the word itself is misspelled, ranging to \textit{data-mining} or \textit{datamining}.

A first definition of \textit{Data Mining} has been introduced by Frawley and Piatetsky-Shapiro \cite{frawley} saying that ``Data Mining is the nontrivial extraction of implicit, previously unknown, and potentially useful information from data''. Since that time, diverse definitions have followed, sometimes slightly deviated, sometimes with the introduction of novel components. One reason for this pragmatic attendance is the domain where the definitive person is associated with; this is regarding the occupational association, the speciality, and the volitional interest. The distribution of databases is always accomplished by an accent of the potential of a Data Warehouse, statisticians surely respond through an importance of mathematical consideration, and research staff from machine learning proclaim Data Mining as the application of Machine Learning per se. However, all of them agree that \textit{Data Mining} is a non-deterministic process, explorative, following a garbage-in and garbage-out paradigm. And the observation from a practical point of view has led to the general belief that different data mining phases exist: firstly, a collection of data, possibly raw, as a response to a previously defined task (originally, this wasn't said, but the complexity had made it inevitable), the pre-processing of the data, the data analysis through algorithms from Machine Learning, partially based on statistical facts and rules, and finally a visualization of the results, matched with an interpretation to understand what has been received. In this respect, commercial and freely available software systems have been created, sometimes praising a \textit{Data Mining-Suite} or \textit{Data Mining-Tool}, paired with the suggestion that the software completely runs autonomously (which is, in fact, untrue, since one important feature of Data Mining is surely the role of the human analyst, being still the definer and decider on almost each levels).

\Section{Alternative Perspective}

Apart from all these histories and the technical side of stringing together data operations, learning algorithms, and visualization techniques, \textit{Data Mining} serves as a much higher impact. Likewise the chemical coherence of gas, fluid, and solid, the importance of \textit{Data Mining} is to transfer a set of data $D$ into an other state of aggregation that allows the user to potentially benefit from it. In other words, \textit{Data Mining} represents a class of digital boosters or catalysts to converse data.

For this, we call the set of data $D$ as the \textit{first state of aggregation}. Whereas data per se simply reflects an increate entity, comparatively an one-dimensional mathematical vector with a value inside, the establishment of information - as a result of techniques described above - proves the existence of something that might be worth or at least interesting. Of course,  information might be redundant as well, but the crux is that information has a meaning, and it must therefore be placed to a different state. In this regard, we call the set of information as the \textit{second state of aggregation} $F$.

\begin{figure}[htbp]
   \centering
   \includegraphics[width=7cm]{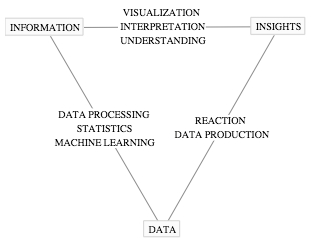} 
   \caption{Three states of aggregation: Data ($D$), Information ($F$), and Insights ($I$).}
   \label{fig:dii1}
\end{figure}

As already mentioned, the discovery or affiliation of information is not always worth, not even for the moment, but obviously or already known. Filtering out this \textit{noise} and concentrating on the immaculate set of information means to hold the nuggets in the hand. These \textit{insights} evolve from an understanding of the situation and a conclusive estimation of what is important and what is not. In this respect, the set of \textit{insights} $I$ represents the \textit{third state of aggregation}, probably the most interesting one, returning to the first state of aggregation $D$ back again. This is done by exploitation of the \textit{insights}, by reacting due to the new situation. To repress $I$ from being exploited simply means to stand still, avoiding the previously defined task to be solved (Figure \ref{fig:dii1}).

But not only a forward but also a backward circulation is available. A return to the second state of aggregation ($I\rightarrow F$) can, for example, be obtained by a verificative check of discovered insights; this possibly lead - as a matter of curiosity or sniffing on novelties - to alternative information, which on his part may be aggregated to $I$. Moreover, changing the aggregation state $F\rightarrow D$ may cause the support of a correction of the data aggregation state itself: maybe, data is false defined, simply wrong or even inconsistent.

A \textit{fourth state of aggregation} $K$ is called as the \textit{knowledge aggregation state}. $K$ potentially exists, but corresponds rather to long-term facts that are true in consequence of affirmed observations. Whereas other states of aggregation are more short-termed and dynamic in some sense, the \textit{knowledge state of aggregation} $K$ is stable and modified rather seldom. Moreover, $K$ occurs continuously, on all states of aggregation.

\begin{figure}[htbp]
   \centering
   \includegraphics[width=7cm]{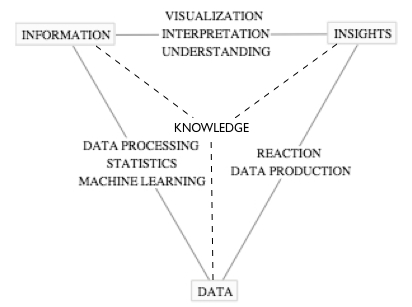} 
   \caption{Introducing the fourth state of aggregation $K$}
   \label{fig:dii2}
\end{figure}

\Section{Consequences}

%
\SubSection{Maintenance and Circulation}
Generally, we have to differ between the \textit{maintenance} of an aggregation state and the transfer to another one (\textit{circulation}). For that, the concepts, techniques, and paradigms that have been mentioned above come into play. For example, a maintenance of data can be achieved by simply managing it through databases, warehouses, or simply data files. The data is given, potentially structured and finite, and continuously be observed, but unstructured data, data streams, or fluent data belong to it as well. Generally, the same principles apply to all states of aggregation: both information, insights and knowledge may be maintained by these solutions, even be enriched by alternative representation formalisms like images, sound, or videos.

The transfer, however, is more especial and not really exchangeable. The transfer of $D\rightarrow F$ is a tedious process, including the data preparation, the non-deterministic usage of statistical and machine learning algorithms, and data visualization as well. With that, information can be extracted, evaluated, and filtered. The transfer $F\rightarrow I$ is solely be done by a (human) interpretation. For example, receiving thousands of association rules may imply thousands of inspections and revisions. The final transfer $I\rightarrow D$ is the conclusion of an aware application.

A final important aspect is - in our understanding - the differentiation between the third and the fourth state of aggregation. The main difference is due to the time, meaning that $I$ corresponds more to a short-termed novelty that is true for the data moment. $K$, however, is independent from the time per se, representing a fact that has been proven again and again. 

\SubSection{Definition}\label{def}
Following the ideas above, our understanding of \textit{Data Mining} is solely concerned with the declaration of states of aggregation. This concerns both the management and the transfer as well, aiming at gaining purely insights (and nothing else, but only if existing): insights stands on his part for novelty through exploration. This can be achieved by verification, which leads to an authentication but not to novelty. In this respect, our understanding of Data Mining is the following:

\begin{itemize}
    \item Data Mining is the maintenance of and circulation in different states of aggregation to finally gain insights.
\end{itemize}

The maintenance concerns both $D$, $F$, $I$, and $K$. As already mentioned, the set of data $D$ can be managed by databases, data warehouse, images, audio files, and videos. The sets of information $F$, insights $I$, and knowledge $K$ are managed by clusters, decision trees, neural networks, a set of association rules, visualizations, statistical values, prediction rules, etc. The circulation is bidirectional and reflexive, meaning that a circulation does not necessarily cause a change of states of aggregation. With this, the circulation includes topics like data pre-processing like discretisation, sampling, factor analysis and principal component analysis, etc., and machine learning algorithms like decision trees, clustering, association discovery, etc.

\begin{figure}[htbp]
   \centering
   \includegraphics[width=7cm]{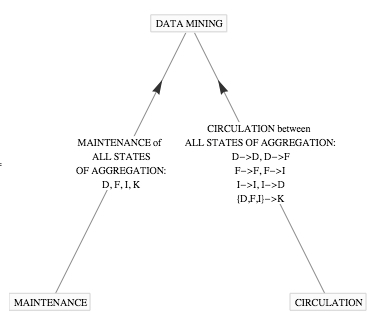} 
   \caption{Data Mining as the symbiosis of Maintenance and Circulation.}
   \label{fig:dii4}
\end{figure}

Note that $I\subseteq F$ because $I$ simply represents the filtered information that is worth and novel. On the other side, $F\subseteq D^n$, having an arbitrary natural number $n$, where the set of information is a subset of the combination of data of arbitrary size. In concern of the dimensions, the absolute value of data elements is not necessarily bigger than the information value. However, $I \neq K$ and $|I|\leq |F|$ because of the filtering again.

\SubSection{Mapping}
We try to map three given definitions exemplarily and therewith prove that the definition given above bears up against this attribution. In this regard, we firstly select a definition given by \cite{fayyad}, then a definition by \cite{wikipedia}, and finally a definition by \cite{kdnuggets}. In \cite{fayyad}, Data Mining is explained as \textit{the non-trivial process of identifying valid, novel, potentially useful, and ultimately understandable patterns in data}, \cite{wikipedia} proposes that Data Mining \textit{is referenced to the process of sorting through large amounts of data and picking out relevant information} and \cite{kdnuggets} argues that Data Mining \textit{is the process of finding new and potentially useful knowledge from data. Or more simpler: Data Mining is the art and science of finding interesting and useful patterns in data.}

For the first definition, the \textit{non-trivial process} certainly reflects the circulation between at least two states of aggregation. The demand of finding patterns of a certain characteristic is ensured as that the third state of aggregation guarantees - through the involvement of the human user - exactly such insights. Sorting through large amounts of data, as mentioned in the second definition, refers to the circulation $D\rightarrow F$ and possibly to $D\rightarrow I$ as well (picking up), although the terminology (information $\neq$ insights) is not congruent. For the third definition, the argument of having a \textit{process} is again the circulation itself whereas the term \textit{knowledge} corresponds to our conception of having \textit{insights} before \textit{knowledge}. The \textit{art and science of} lies in the hands of the human user, which is also given here.

\SubSection{Example}
Let $D$ the set of data, $F$ the set of information, and $I$ the set of insights, and assume that the following simple data situation is given with a number of web users and a number of web pages being visited. Assume furthermore that only web page names are stored, but not the suffix \textit{.html}. The task is to analyse the users' web behavior, possibly leading to a re-organisation of the existing web portal structure. For that, the standard association discovery (apriori) (\cite{agrawal}) is applied, setting initially the threshold parameters to \textit{support=2} and \textit{confidence=2}.

{\small
\begin{verbatim}
D = {
User 1 : home, baby, drinks, wclothes
User 2 : music, cclothes, wclothes,
         cclothes, drinks, wclothes
User 3 : baby, drinks, baby, info, home 
User 4 : drinks, home, fashion, wclothes 
User 5 : baby, drinks, wclothes, fashion,
         cclothes, wclothes
User 6 : home, info, wclothes, cclothes,
         fashion, wclothes 
...
}
\end{verbatim}
}
The data situation corresponds to the first state of aggregation that follows an reflexive aggregation procedure of extracting the web page names out of the log-file, the fixation of the threshold parameters and the structuring of the data per se. In principle, this can be extended to the establishment of a hierarchy, for example that clothes for women (wclothes), men (mclothes), and children (cclothes) belong to the superior class \textit{clothes}.

The first state of aggregation becomes abandoned when calculating a set of association rules, defining the second state of aggregation. Furthermore, several other information can be gained by observation, for example that \textit{wclothes} is often used as exit or that \textit{home} is rather a rare entrance page. The information aggregation state therefore is 

{\small
\begin{verbatim}
F = {
...
fashion ==> wclothes
baby ==> drinks
...
wclothes: exit page in 80% of cases 
home: entrance page in 33% of cases.
}
\end{verbatim}
}
with $F\subset D^n$. In fact, the association rules are accomplished by statistical values to ease the interpretation; at this moment, the second state of aggregation changes to the third state of aggregation ($F\rightarrow I$). The change to the insights state of aggregation is done by interpretation: whereas \textit{fashion} $\rightarrow$ \textit{wclothes} is obvious and filtered out, only the fact concerning \textit{baby} and \textit{drinks} as well as the \textit{exit} story, respectively, is taken:

{\small
\begin{verbatim}
I = {
baby ==> drinks
wclothes is often used as exit 
}
\end{verbatim}
}
With that, and as a consequence of $I\rightarrow D$, the web page system may be restructured, offering a separate link on each \textit{baby} web-page to \textit{drinks}. In the future, this slightly change will influence the data state of aggregation. Furthermore, the fact that a visit of \textit{baby.html} triggers a visit of \textit{drinks.html} certainly may invite to verify whether other beverages are also concerned or not. The worth of the \textit{baby, drinks} correlation may likewise incite to additionally consider hierarchies or ontologies. In fact, there are many reasons to branch backward to states of aggregation.

To get back to the differentiation between the third and fourth state of aggregation: the fact that a correlation between \textit{baby} and \textit{drinks} exists is certainly to be put on the insight state of aggregation but not necessarily on the knowledge state of aggregation. However, if it becomes certified that - for example - unconcerned sets of data show a similar insight or that \textit{baby}$\rightarrow$\textit{drinks} is present in similar web systems as well, then this insight is definitely circulated to $K$.

\SubSection{Criticism}
We have defined the term \textit{Data Mining} consciously as it has been introduced in section \ref{def}, and with this, it consequently affects long-established research fields like database systems, statistics, and Artificial Intelligence. In this respect, the definition, however, could be too comprehensive to someone's position and/or probably overdraw a general accordance and acceptance. This is intelligible but everyone should be clear in his mind that a discussion about \textit{Data Mining} is per se not of small value and that a consideration of influencing techniques must be considered. As an example, the field of \textit{Artificial Intelligence} also uses, among other disciplines, logic, linguistic ideas, or programming languages: it is therefore not superior but simply understands these research fields as legitimate additives.

\Section{Conclusions}

To reconcile existing derivatives of \textit{Data Mining} and to discover a definition for it is necessary and a matter of course. The current situation in the field of computer science and its interdisciplinary fields do not follow this demand but pragmatically take absorb existing explanations or devise novel topics in it. This work surely contributes to this as well, but tries to constructively propose a uniform prospect of \textit{Data Mining}. Following our experiences in industry and academics, we believe that the proposed model of states of aggregation in conjunction with the maintenance and circulation process is a reliable, confident and constructive model.

\section*{Acknowledgement}
This work has been done in the scope of the research project ICC, which is currently performed at the MINE research group, ILIAS Computer Science Laboratory. 


\end{document}